\documentclass[aps,twocolumn,preprintnumbers,amsmath,amssymb]{revtex4}

\usepackage{graphicx}
\usepackage{dcolumn}
\usepackage{bm}
\usepackage{color}

\begin{document}


\title{Localized Higgs modes of superfluid Bose gases in optical lattices: 
\\ A Guzwiller mean-field study}
\author{Ippei Danshita$^{1}$}\email{danshita@yukawa.kyoto-u.ac.jp}
\author{Shunji Tsuchiya$^{2}$}\email{tsuchiya@phys.chuo-u.ac.jp}
\affiliation{
{$^1$Yukawa Institute for Theoretical Physics, Kyoto University, Kitashirakawa Oiwakecho, Sakyo-ku, Kyoto 606-8502, Japan}
\\
{$^2$Department of Physics, Chuo University, 1-13-27 Kasuga, Bunkyo-ku, Tokyo 112-8551, Japan}
}

\date{\today}

\begin{abstract}
We study effects of a potential barrier on collective modes of superfluid Bose gases in optical lattices. We assume that the barrier is created by local suppression of the hopping amplitude. When the system is in a close vicinity of the Mott transition at commensurate fillings, where an approximate particle-hole symmetry emerges, there exist bound states of Higgs amplitude mode that are localized around the barrier. By applying the Gutzwiller mean-field approximation to the Bose-Hubbard model, we analyze properties of normal modes of the system with a special focus on the Higgs bound states. We show that when the system becomes away from the Mott transition point, the Higgs bound states turn into quasi-bound states due to inevitable breaking of the particle-hole symmetry. We use a stabilization method to compute the resonance energy and line width of the quasi-bound states. 
We compare the results obtained by the Gutzwiller approach with those by the Ginzburg-Landau theory. We find that the Higgs bound states survive even in a parameter region far from the Mott transition, where the Ginzburg-Landau theory fails. 

\end{abstract}


\maketitle
\section{Introduction}
A Higgs mode is a gapful collective mode that exists generally in a system with approximate particle-hole symmetry and spontaneous breaking of continuous symmetry, and it corresponds to fluctuations of the amplitude of an order parameter~\cite{volovik-14,pekker-15}. Recently, this type of collective mode has attracted much attention thanks to experimental observations in various condensed-matter and quantum-gas systems, such as superconductors ${\rm NbSe}_2$~\cite{sooryakumar-80,littlewood-81,measson-14} and ${\rm Nb}_{1-x}{\rm Ti}_x{\rm N}$~\cite{matsunaga-13,matsunaga-14,sherman-15,matsunaga-17}, quantum antiferromagnets ${\rm TlCuCl}_3$~\cite{ruegg-08,merchant-14} and ${\rm KCuCl}_3$~\cite{kuroe-12}, charge-density-wave materials ${\rm K}_{0.3}{\rm MoO}_3$~\cite{demsar-99,schaefer-14} and ${\rm TbTe}_3$~\cite{yusupov-10,mertelj-13}, superfluid ${}^3{\rm He}$ B-phase~\cite{avenel-80,collett-13}, and superfluid Bose gases in optical lattices~\cite{bissbort-11,endres-12}.

While most of previous studies on Higgs modes have regarded states delocalized in an entire system, the current authors and coworkers have recently pointed out that in the presence of a potential barrier there also exist Higgs modes localized around the barrier whose excitation energy is below the gap of the delocalized Higgs mode in a bulk~\cite{nakayama-15}. 
They derived the fourth-order Ginzburg-Landau (GL) theory from the Bose-Hubbard model describing an ultracold Bose gas in an optical lattice and used it to find the Higgs bound states in the superfluid phase. However, the parameter region in which the GL theory is valid is limited to a close vicinity of the Mott transition so that experimental observation of the Higgs bound states will require rather fine tuning of parameters if they exist only in the validity region of the GL theory. Although such fine tuning is available with recent experimental technology, it is better if the Higgs bound states exist in a broader parameter region. Moreover, the excitation energies of the Higgs bound states are found to be very close to the bulk Higgs gap in the GL analysis. Since a collective mode is usually detected in experiment as a resonance peak in response to external perturbations and such a peak is broadened to some extent by quantum and thermal fluctuations, this property may prevent a Higgs bound state from being observed separately from the delocalized Higgs mode. 

In the present paper, we analyze the Higgs bound states of superfluid Bose gases in optical lattices by applying the Gutzwiller mean-field approximation to the Bose-Hubbard model with local suppression of the hopping amplitude. The Gutzwiller approach allows us to explore the Higgs bound states in a broader parameter region including the one far from the validity region of the GL theory. The particle-hole symmetry of the system is only approximate except at the Mott transition points at commensurate fillings, and such slight breaking of the particle-hole symmetry couples the Higgs bound states with delocalized Nambu-Goldstone (NG) modes, converting the former into quasi-bound states with finite life time. In order to obtain the resonance energy and line width of quasi-bound states, we use a stabilization method developed by Mandelshtam {\it et al.}~\cite{mandelshtam-93}, in which the density of states for a quasi-bound state is constructed from the system-size dependence of the excitation energies. 

Using the methods mentioned above, we investigate how properties of the Higgs bound states depend on the global hopping amplitude, the chemical potential, and the spatial width of the local hopping suppression. We show the presence of the Higgs bound states even in the extended parameter region where the GL theory is invalid. We also show that the number of the bound states increases with the width. Moreover, we find that by tuning the width and the global hopping amplitude optimally the energy of the lowest Higgs bound state can be significantly separated from those of other collective modes in comparison with the case of the GL prediction. Such large energy separation is advantageous for observing the lowest Higgs bound state in experiment.

The remainder of the paper is organized as follows. In Sec.~\ref{sec:model}, we introduce the Bose-Hubbard model and explain our potential barrier created by local suppression of the hopping amplitude. In Sec.~\ref{sec:method}, we review the GL theory derived from the Bose-Hubbard model in the presence of a potential barrier and solutions of the GL equation corresponding to the Higgs bound states. We also review a numerical method to analyze normal modes of the Bose-Hubbard model within the Gutzwiller mean-field approximation. Moreover, we describe how to utilize a stabilization method in our discrete (lattice) system. In Sec.~\ref{sec:result}, varying the global hopping amplitude, the chemical potential, and the barrier width, we analyze properties of the Higgs bound states. In Sec.~\ref{sec:conc}, we summarize the results.

\section{Model}
\label{sec:model}

We consider a system of a Bose gas in a hypercubic optical lattice with lattice spacing $a$ at zero temperature. We assume that the optical-lattice potential is sufficiently deep so that the system is well described by the Bose-Hubbard model~\cite{fisher-89,jaksch-98},
\begin{eqnarray}
\hat{H} = -\sum_{{\boldsymbol i},{\boldsymbol j}} J_{{\boldsymbol i},{\boldsymbol j}}
\hat{b}_{\boldsymbol i}^{\dagger} \hat{b}_{\boldsymbol j}
- \sum_{\boldsymbol i} \mu_{\boldsymbol i} \hat{b}_{\boldsymbol i}^{\dagger} \hat{b}_{\boldsymbol i}
+\frac{U}{2}\sum_{\boldsymbol i} \hat{b}_{\boldsymbol i}^{\dagger}\hat{b}_{\boldsymbol i}^{\dagger} \hat{b}_{\boldsymbol i} \hat{b}_{\boldsymbol i},
\label{eq:BHM}
\end{eqnarray}
where $\hat{b}_{\boldsymbol i}$ ($\hat{b}_{\boldsymbol i}^{\dagger}$) annihilates (creates) a boson at site ${\boldsymbol i}$.
The vector ${\boldsymbol i}\equiv \sum_{\alpha = 1}^d i_{\alpha} {\boldsymbol e}_{\alpha}$ denotes the site index, where $i_{\alpha}$ is an integer, $d$ the spatial dimension, and ${\boldsymbol e}_{\alpha}$ a unit vector in direction $\alpha$.  $U$$(>0)$ and $\mu_{\boldsymbol i}$ represent the onsite repulsion and the local chemical potential.
As the hopping matrix element, we assume the following nearest-neighbor form,
\begin{eqnarray}
J_{{\boldsymbol i},{\boldsymbol j}} = 
\sum_{\alpha}\left( 
J_{\boldsymbol j}^{(\alpha)}
\delta_{{\boldsymbol i},{\boldsymbol j}+{\boldsymbol e}_{\alpha}} 
+
J_{{\boldsymbol j}-{\boldsymbol e}_{\alpha}}^{(\alpha)}
\delta_{{\boldsymbol i},{\boldsymbol j}-{\boldsymbol e}_{\alpha}} 
\right),
\end{eqnarray}
where $J_{\boldsymbol j}^{(\alpha)}$ means the hopping amplitude between sites ${\boldsymbol j}$ and ${\boldsymbol j}+{\boldsymbol e}_{\alpha}$.
We set $\hbar = a = 1$ throughout the paper except in Figures.

While the purpose in the present paper is to discuss the effect of inhomogeneity in $J_{\boldsymbol j}^{(\alpha)}$ on collective modes in the superfluid phase, here we briefly review properties of ground states and collective modes of the Bose-Hubbard model (\ref{eq:BHM}) in the case that the system is homogeneous ($J_{\boldsymbol j}^{(\alpha)} = J$, $\mu_{\boldsymbol j}=\mu$). When the filling factor $n$ (the number of atoms per site) is non-integer, the ground state is always a superfluid phase. In contrast, when $n$ is integer, namely at commensurate fillings, and $zJ/U$ is decreased from the superfluid phase, the quantum phase transition to a Mott insulating phase occurs~\cite{fisher-89}, where $z$ denotes the coordination number. The superfluid-to-Mott insulator transitions have been experimentally observed by tuning the optical-lattice depth, corresponding to the control of the ratio $zJ/U$~\cite{greiner-02}.

The Mott insulating phase has particle and hole excitations~\cite{elstner-99,vanoosten-01,konabe-06}.
The finite energy gap in these excitations reflects the insulating nature of the phase.
In general, the superfluid phase, in which a global U(1) symmetry of the system is spontaneously broken, has a gapless NG mode corresponding to fluctuations of the phase of the superfluid order-parameter. In a vicinity of the Mott transition at commensurate fillings, an effective particle-hole symmetry emerges such that there exists an additional collective mode, namely a gapful Higgs mode, corresponding to amplitude fluctuations~\cite{altman-02,huber-07}.

\begin{figure}[tb]
     \includegraphics[scale=0.52]{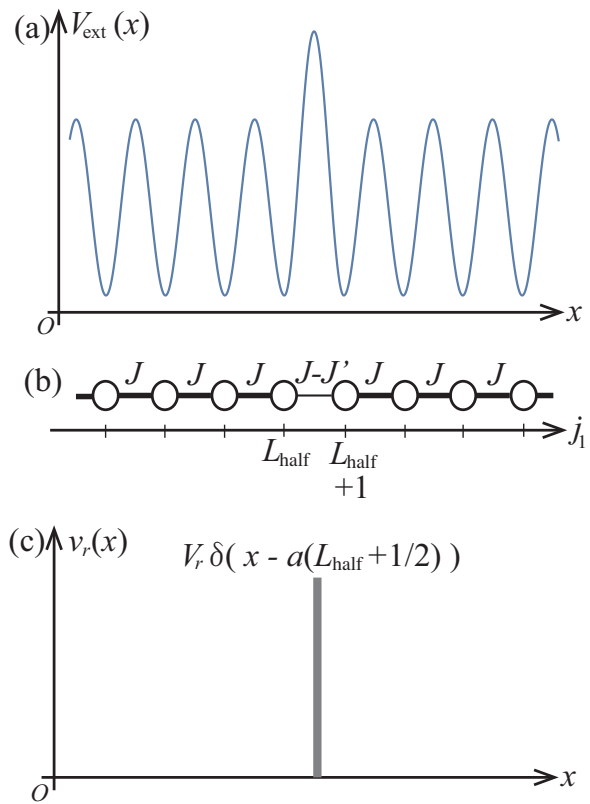}
   \caption{Schematic illustration of a potential barrier created by local suppression of the hopping energy for the case of $w=0$. (a) The external potential for atoms, which consists of a periodic optical lattice potential and a focused repulsive potential on a single link, is shown. (b) The lattice sites connected via nearest-neighbor hopping is shown, where the hopping is locally suppressed at the link between $L_{\rm half}$ and $L_{\rm half}+1$ sites. (c) In the GL equation, the hopping suppression can be approximated as a delta-functional repulsive barrier.
   }
\label{fig:hop-bar}
\end{figure} 

In Ref.~\cite{nakayama-15}, the current authors and coworkers studied effects of barrier potentials on the collective modes in the superfluid phase. Especially, they considered local suppression of the hopping amplitude in the following form,
\begin{eqnarray}
J_{\boldsymbol j}^{(\alpha)} = J - J_{j_1}'\delta_{\alpha,1},
\end{eqnarray}
and found bound states of the Higgs mode within the GL theory, which is valid only near the Mott transition. In the present work, we analyze such Higgs bound states in greater details by employing the Gutzwiller variational approximation, which allows for the numerical analyses of the collective modes even outside the validity region of the GL theory. Specifically, we consider the hopping suppression of the step-function shape,
\begin{eqnarray}
J_{j_1}' = \left\{
\begin{array}{cc}
J', & {\rm when}\,\, |j_1 - L_{\rm half}| \leq w,
\\
0, & {\rm otherwise},
\end{array}
\right.
\end{eqnarray}
where
\begin{eqnarray}
L_{\rm half} = \left\{
\begin{array}{cc}
L/2, & {\rm when}\,\, L \in {\rm even},
\\
(L+1)/2, & {\rm when}\,\, L \in {\rm odd}.
\end{array}
\right.
\end{eqnarray}
$J'$ and $2w+1$ represent the height and width of the hopping suppression. $L$ denotes the number of lattice sites in the $x$-direction. Figures~\ref{fig:hop-bar}(a) and \ref{fig:hop-bar}(b) illustrate the external potential for creating the hopping suppression and the spatial dependence of the hopping amplitude in the $x$-direction in the case of $w=0$. Such local control of the external potential is achievable, e.g., with use of a digital micro-mirror device~\cite{islam-15}.

In the previous work~\cite{nakayama-15}, the effects of local spatial modulation in the chemical potential $\mu_{\boldsymbol i}$, which explicitly breaks the particle-hole symmetry, have been also considered. As a consequence, it has been shown that the NG mode incident to the potential barriers exhibits a Fano resonant tunneling. However, in the present paper, we assume that the chemical potential is homogeneous, i.e. $\mu_{\boldsymbol i}=\mu$, and focus only on the potential barrier created by the local suppression of the hopping amplitude, which does not break the particle-hole symmetry.

\section{Methods}
\label{sec:method}
In this section, we describe two methods for analyzing collective modes of superfluid Bose gases in optical lattices. The first method, namely the GL theory, was used for finding the Higgs bound states in Ref.~\cite{nakayama-15} and is briefly reviewed in Sec.~\ref{subsec:GL}. While it provides us with useful analytical solutions of the Higgs bound states, the parameter region in which it is applicable is limited only to a vicinity of the Mott transition. The second method is a mean-field approximation based on the Gutzwiller variational wave function and is explained in Sec.~\ref{subsec:Gutz}. In contrast to the GL theory, the latter is fully numerical and can capture properties of the collective modes even in a region far away from the quantum phase transition. In this sense, the two methods are complementary to each other. 

Except at the quantum critical point, the Higgs bound states are coupled with NG modes to some extent because of the higher order terms in the GL expansion that break particle-hole symmetry while such couplings are neglected in the analytical solutions of the GL equation. Due to the couplings the Higgs bound states can turn into a quasi-bound state with a finite life time. In order to accurately compute the resonance energy and the life time of the quasi-bound states, we combine the Gutzwiller approach with a stabilization method developed by Mandelshtam {\it et al.}~\cite{mandelshtam-93}, as shown in Sec.~\ref{subsec:SM}.

It is well known that when $d<3$, the two methods, which neglect effects of quantum fluctuations from the mean fields, fail to correctly describe collective modes of the superfluid phase near the Mott transition at commensurate fillings. Specifically, at the low dimensions, such fluctuations enhance the decay channel of a Higgs mode into two NG modes so strongly that the Higgs mode is not necessarily well defined~\cite{altman-02,sachdev-99,podolsky-11,pollet-12,gazit-13,chen-13,rancon-14,liu-15,katan-15,rose-15}. Hence, in the following, we only consider the case of $d=3$, where the Higgs mode is known to be long-lived~\cite{affleck-92,altman-02,nagao-16} so that the use of the two methods is well justified.
We also note that in the Gutzwiller calculations shown below we assume a periodic boundary condition in the $x$-direction and homogeneity in the other directions.
\subsection{Ginzburg-Landau theory}
\label{subsec:GL}
When the system is in the superfluid phase in a vicinity of the Mott transition, the amplitude of the superfluid order-parameter is so small that an effective GL action describing the dynamics of the superfluid order-parameter $\Phi({\boldsymbol x},t)$ can be derived~\cite{nakayama-15,sachdev-11,kato-14} through a perturbative expansion. Taking the saddle-point approximation of the action with respect to $\Phi^\ast({\boldsymbol x},t)$ leads to the time-dependent GL equation,
\begin{eqnarray}
iK_0 \frac{\partial \Phi}{\partial t} \! -\! W_0 \frac{\partial^2 \Phi}{\partial t^2}
\!=\! \left(\!
-\frac{\nabla^2}{2m_{\ast}}+ r_0 + v_r(x) + u_0 |\Phi|^2 
\!\right)\! \Phi,
\label{eq:GL}
\end{eqnarray}
where the coefficients $K_0$, $W_0$, $m_{\ast}$, $r_0$, $v_r(x)$, and $u_0$ are explicitly related to the parameters $J_{{\boldsymbol i},{\boldsymbol j}}$, $U$, and $\mu$ in the original Bose-Hubbard model (\ref{eq:BHM}). At commensurate fillings, $K_0\simeq 0$ such that the GL equation (\ref{eq:GL}) is particle-hole symmetric, i.e., it is symmetric with respect to replacing $\Phi$ with $\Phi^{\ast}$. As will be shown below, the condition $K_0=0$ is necessary for the Higgs amplitude mode to exist as an independent collective mode.

In order to simplify the notation, we rewrite the variables in a dimensionless form as follows:
\begin{eqnarray}
\bar{\Phi} \!\!&=&\!\! \Phi/(-r_0/u_0)^{1/2}, \,\, \bar{t}=t(-r_0/W_0)^{1/2}, \,\, \bar{x}=x/\xi,
\nonumber \\
 \bar{v}_r \!\!&=&\!\! v_r/(-r_0), \,\, \bar{K}_0=K_0/(-r_0W_0)^{1/2},
\end{eqnarray}
where $\xi = (-r_0m_{\ast})^{-1/2}$ denotes the healing length.
In this way, the time-dependent GL is converted to a dimensionless form,
\begin{eqnarray}
iK_0 \frac{\partial \Phi}{\partial t} - \frac{\partial^2 \Phi}{\partial t^2}
= \left(
-\frac{\nabla^2}{2} - 1 + v_r(x) + |\Phi|^2 
\right) \Phi ,
\label{eq:GL_dl}
\end{eqnarray}
where we omitted the bars on the variables for simplicity.

In order to describe collective modes, we separate the superfluid order-parameter into its static value and small fluctuations,
\begin{eqnarray}
\Phi({\boldsymbol x},t) &=& \tilde{\Phi}({\boldsymbol x}) 
+ \left[S({\boldsymbol x}) + T({\boldsymbol x})\right]e^{-i\omega t} 
\nonumber \\
&& + \left[S^{\ast}({\boldsymbol x}) - T^{\ast}({\boldsymbol x})\right]e^{i\omega^{\ast} t}.
\end{eqnarray}
The static part of the order parameter $\tilde{\Phi}({\boldsymbol x})$ obeys the following nonlinear equation,
\begin{eqnarray}
\left[
-\frac{\nabla^2}{2} - 1 + |\tilde{\Phi}({\boldsymbol x})|^2 + v_r(x)
\right]\tilde{\Phi}({\boldsymbol x}),
\label{eq:GP}
\end{eqnarray}
which is identical to the static Gross-Pitaevskii equation~\cite{pitaevskii-03}.
In the fluctuation parts, $S({\boldsymbol x})$ and $T({\boldsymbol x})$ correspond to phase and amplitude fluctuations, respectively, and they obey a set of linear equations,
\begin{eqnarray}
\left[
-\frac{\nabla^2}{2} - 1 + |\tilde{\Phi}({\boldsymbol x})|^2 + v_r(x)
 \right]
S({\boldsymbol x}) 
\nonumber \\ 
\!\!\!\!\!\!\!\!\!\!\!\!\!\!\!\!\!\!\!\!\!\!\!\!\!\!\!\!\!\!\!\!\!\!\!\!\!\!\!\! 
= \omega^2 S({\boldsymbol x}) + K_0 \omega T({\boldsymbol x}),
\\
\left[
-\frac{\nabla^2}{2} - 1 + 3|\tilde{\Phi}({\boldsymbol x})|^2  +  v_r(x)
\right]T({\boldsymbol x}) 
\nonumber \\
\!\!\!\!\!\!\!\!\!\!\!\!\!\!\!\!\!\!\!\!\!\!\!\!\!\!\!\!\!\!\!\!\!\!\!\!\!\!\!\!  
= \omega^2 T({\boldsymbol x}) + K_0 \omega S({\boldsymbol x}).
\label{eq:Higgs}
\end{eqnarray}
When the system is particle-hole symmetric, i.e., $K_0 = 0$, the phase and amplitude fluctuations are decoupled from each other so that the Higgs mode may exist independently in addition to the NG mode. In the absence of the potential barrier, the dispersions of the two modes are given by
\begin{eqnarray}
\begin{array}{cc}
{\rm NG:}\,\, \omega^2 = c^2 k^2, \\
{\rm Higgs:}\,\, \omega^2 = c^2 k^2 + \Delta^2,
\end{array}
\end{eqnarray}
where $c=2^{-1/2}$ and $\Delta = 2^{1/2}$ denote the sound speed and the Higgs gap in the bulk. 
 An approximate particle-hole symmetry indeed emerges in a vicinity of the Mott transition points at commensurate fillings~\cite{sachdev-11}. In the remainder of this section, we assume $K_0 = 0$ and describe the analytical solution of a Higgs bound state obtained in Ref.~\cite{nakayama-15}.
 
We assume that the healing length $\xi$ is sufficiently large compared to the width of the hopping suppression $2w+1$. In this situation, the potential barrier in the GL equation can be approximated as a $\delta$-function form,
\begin{eqnarray}
v_r(x) = V_r \delta\left(x - L_{\rm half}+1/2\right).
\end{eqnarray}
When $J' \ll J$, $V_r$ can be approximately related to $J'$ as
\begin{eqnarray}
V_r = 2J'(2w+1).
\end{eqnarray}
The potential barrier is schematically depicted in Figure~\ref{fig:hop-bar}(c).
For simplicity, we hereafter shift the location of the potential barrier to the origin of our frame, i.e., we set
\begin{eqnarray}
v_r(x) = V_r \delta\left(x\right).
\label{eq:barrier_GL}
\end{eqnarray}

The analytical solution of the static equation (\ref{eq:GP}) with the potential barrier of Eq.~(\ref{eq:barrier_GL}) is given by~\cite{kovrizhin-01}
\begin{eqnarray}
\tilde{\Phi}(x) = \tanh\left(|x|+x_0\right),
\label{eq:tanh}
\end{eqnarray}
where
\begin{eqnarray}
\tanh\left(x_0\right) = -\frac{V_r}{2} + \sqrt{\frac{V_r^2}{4}+1}.
\end{eqnarray}
Substituting Eq.~(\ref{eq:tanh}) into Eq.~(\ref{eq:Higgs}) and seeking solutions of the latter with $\omega < \Delta$, we obtain bound-state solutions of the Higgs mode~\cite{nakayama-15}. Here we explicitly write the one with even parity,
\begin{eqnarray}
T(x) = \left( 3\tilde{\Phi}^2 + 3 \kappa_+ \tilde{\Phi} + \kappa_+^2 -1 \right) e^{-\kappa_+ |x|},
\label{eq:Tx}
\end{eqnarray}
where $\kappa_+ = \sqrt{4-2E_+^2}$.
The binding energy $E_+$ is determined as the solution of the boundary condition at $x=0$,
\begin{eqnarray}
\left.\frac{dT}{dx}\right|_{x=+0} - \left.\frac{dT}{dx}\right|_{x=+0} = 2V_r T(0).
\end{eqnarray}
This even Higgs bound state always exists when $V_r > 0$. Another Higgs bound state, which has odd parity, emerges when the barrier strength exceeds a certain threshold value~\cite{nakayama-15}. The emergence of these Higgs bound states can be attributed to the formation of an effective double well potential for $T(x)$ created by the combination of $V_r\delta(x)$ and $3\tilde{\Phi}(x)^2$. Notice that the same bound-state solutions of the GL equation were discussed also in the context of superconductors~\cite{hammer-16}.
In Sec.~\ref{subsec:comp}, we compare the analytical solutions (\ref{eq:tanh}) and (\ref{eq:Tx}) directly with the results by the Gutzwiller mean-field theory.

\subsection{Gutzwiller mean-field approximation}
\label{subsec:Gutz}
In the Gutzwiller mean-field approximation, the many-body wave function of the system is approximated by the following variational wave function~\cite{rokhsar-91},
\begin{eqnarray}
|\Psi_{\rm GW} \rangle = \prod_{\boldsymbol i}
 \sum_{n} f_{{\boldsymbol i},n}(t)|n\rangle_{\boldsymbol i},
\end{eqnarray}
which forms a single product of local states. $|n\rangle_{\boldsymbol i}$ represents the local Fock state at site ${\boldsymbol i}$. The variational parameter $f_{{\boldsymbol i},n}$ satisfies the normalization condition,
\begin{eqnarray}
\sum_n |f_{{\boldsymbol i},n}|^2 = 1,
\label{eq:norm_f}
\end{eqnarray}
and the equation of motion,
\begin{eqnarray}
i\frac{d}{dt}f_{{\boldsymbol i},n} &=& 
- \sum_{\boldsymbol j} J_{{\boldsymbol i},{\boldsymbol j}}
\left[
\sqrt{n}f_{{\boldsymbol i},n-1}\Phi_{\boldsymbol j}
+ \sqrt{n+1}f_{{\boldsymbol i},n+1}\Phi_{\boldsymbol j}^{\ast}
\right]
\nonumber \\
&&
+ \left[\frac{U}{2}n(n-1) - \mu_{\boldsymbol i} n
\right]f_{{\boldsymbol i},n},
\label{eq:EOM_f}
\end{eqnarray}
where the superfluid order-parameter is given by
\begin{eqnarray}
\Phi_{\boldsymbol i}\equiv \langle \Psi_{\rm GW}|\hat{b}_{\boldsymbol i} |\Psi_{\rm GW}\rangle = \sum_{n}\sqrt{n}f_{{\boldsymbol i},n-1}^{\ast}f_{{\boldsymbol i},n}.
\label{eq:phi_f}
\end{eqnarray}
While the Gutzwiller approach is a simple mean-field approximation, it has been extensively applied to describing various properties of the Bose-Hubbard model, such as quantum phase transitions~\cite{sheshadri-93,kovrizhin-05,iskin-11}, collective modes~\cite{kovrizhin-05,kovrizhin-07,krutitsky-11,saito-12}, the superfluid critical momentum~\cite{saito-12,altman-05}, and non-equilibrium dynamics~\cite{saito-12,altman-05,snoek-07,krutitsky-10,bissbort-11,inaba-16}. 

In order to describe collective modes of the system, we separate $f_{{\boldsymbol i},n}(t)$ into its static part and small fluctuations,
\begin{eqnarray}
f_{{\boldsymbol i},n}(t) = 
\left[\tilde{f}_{{\boldsymbol i},n} + \delta f_{{\boldsymbol i},n}(t)\right]
e^{-i\tilde{\omega}_{\boldsymbol i}t},
\label{eq:fluc_f}
\end{eqnarray}
where
\begin{eqnarray}
\delta f_{{\boldsymbol i},n}(t) = 
u_{{\boldsymbol i},n}e^{-i\omega t}
+ v_{{\boldsymbol i},n}^{\ast}e^{i\omega^{\ast}t}.
\end{eqnarray}
The static part $\tilde{f}_{{\boldsymbol i},n}$ and the phase factor $\tilde{\omega}_{\boldsymbol i}$ obey
\begin{eqnarray}
\tilde{\omega}_{\boldsymbol i} &=& 
-\sum_{\boldsymbol j} J_{{\boldsymbol i},{\boldsymbol j}}
\left(
\tilde{\Phi}_{\boldsymbol i}^{\ast} \tilde{\Phi}_{\boldsymbol j} + {\rm c.c.}
\right)
\nonumber \\ 
&&
+ \sum_{n} \left[ \frac{U}{2}n(n-1) - \mu_{\boldsymbol i}n \right]|
\tilde{f}_{{\boldsymbol i},n}|^2,
\label{eq:static_f}
\end{eqnarray}
where $\tilde{\Phi}_{\boldsymbol i}= \sum_{n}\sqrt{n}\tilde{f}_{{\boldsymbol i},n-1}^{\ast}\tilde{f}_{{\boldsymbol i},n}$.
Solving Eqs.~(\ref{eq:static_f}) and (\ref{eq:norm_f}) simultaneously, one obtains $\tilde{f}_{{\boldsymbol i},n}$ and $\tilde{\omega}_{\boldsymbol i}$. If the ground state is only of interest among the static solutions, one may alternatively solve Eq.~(\ref{eq:EOM_f}) in imaginary time~\cite{yamashita-07} to obtain $\tilde{f}_{{\boldsymbol i},n}$ and substitute it into Eq.~(\ref{eq:static_f}) to obtain $\tilde{\omega}_{\boldsymbol i}$. We use the latter method for the calculations shown below.

Linearizing the equation of motion (\ref{eq:EOM_f}) with respect to the fluctuations, one obtains a set of linear equations,
\begin{widetext}
\begin{eqnarray}
\omega u_{{\boldsymbol i},n} &=& - \sum_{m}\sum_{\boldsymbol j}
J_{{\boldsymbol i},{\boldsymbol j}}
\left[
\left\{
\sqrt{nm}\tilde{f}_{{\boldsymbol i},n-1}
\tilde{f}_{{\boldsymbol j},m-1}^{\ast}
+ \sqrt{(n+1)(m+1)}\tilde{f}_{{\boldsymbol i},n+1}
\tilde{f}_{{\boldsymbol j},m+1}^{\ast}
\right\} u_{{\boldsymbol j},m}
\right.
\nonumber \\
&& 
\,\,\,\,\,\,\,\,\,\,\,\,\,\,\,\,\,\,\,\,\,\,\,\,\,\,\,\,\,
\left.
+ \left\{
\sqrt{n(m+1)}\tilde{f}_{{\boldsymbol i},n-1}
\tilde{f}_{{\boldsymbol j},m+1}
+ \sqrt{(n+1)m}\tilde{f}_{{\boldsymbol i},n+1}
\tilde{f}_{{\boldsymbol j},m-1}
\right\}v_{{\boldsymbol j},m}
\right]
\nonumber \\
&& 
- \sum_{\boldsymbol j}J_{{\boldsymbol i},{\boldsymbol j}}
\left[
\sqrt{n}\tilde{\Phi}_{\boldsymbol j} u_{{\boldsymbol i},n-1}
+ \sqrt{n+1}\tilde{\Phi}_{\boldsymbol j}^{\ast} u_{{\boldsymbol i},n+1}
\right]
+ \left[ \frac{U}{2}n(n-1) - \mu_{\boldsymbol i}n - \tilde{\omega}_{\boldsymbol i} \right]
u_{{\boldsymbol i},n},
\label{eq:bogo_u}
\\
-\omega v_{{\boldsymbol i},n} &=& - \sum_{m}\sum_{\boldsymbol j}
J_{{\boldsymbol i},{\boldsymbol j}}
\left[
\left\{
\sqrt{nm}\tilde{f}_{{\boldsymbol i},n-1}^{\ast}
\tilde{f}_{{\boldsymbol j},m-1}
+ \sqrt{(n+1)(m+1)}\tilde{f}_{{\boldsymbol i},n+1}^{\ast}
\tilde{f}_{{\boldsymbol j},m+1}
\right\} v_{{\boldsymbol j},m}
\right.
\nonumber \\
&& 
\,\,\,\,\,\,\,\,\,\,\,\,\,\,\,\,\,\,\,\,\,\,\,\,\,\,\,\,\,
\left.
+ \left\{
\sqrt{n(m+1)}\tilde{f}_{{\boldsymbol i},n-1}^{\ast}
\tilde{f}_{{\boldsymbol j},m+1}^{\ast}
+ \sqrt{(n+1)m}\tilde{f}_{{\boldsymbol i},n+1}^{\ast}
\tilde{f}_{{\boldsymbol j},m-1}^{\ast}
\right\}u_{{\boldsymbol j},m}
\right]
\nonumber \\
&& 
- \sum_{\boldsymbol j}J_{{\boldsymbol i},{\boldsymbol j}}
\left[
\sqrt{n}\tilde{\Phi}_{\boldsymbol j}^{\ast} v_{{\boldsymbol i},n-1}
+ \sqrt{n+1}\tilde{\Phi}_{\boldsymbol j} v_{{\boldsymbol i},n+1}
\right]
+ \left[ \frac{U}{2}n(n-1) - \mu_{\boldsymbol i}n - \tilde{\omega}_{\boldsymbol i} \right]
v_{{\boldsymbol i},n},
\label{eq:bogo_v}
\end{eqnarray}
\end{widetext}
which determines the wave function $(u_{{\boldsymbol i},n},v_{{\boldsymbol i},n})$ and the frequency $\omega$ of normal modes. We assume that $(u_{{\boldsymbol i},n},v_{{\boldsymbol i},n})$ satisfies the normalization condition,
\begin{eqnarray}
\sum_{\boldsymbol i}\sum_{n}
\left( \left|u_{{\boldsymbol i},n}\right|^2 - \left|v_{{\boldsymbol i},n}\right|^2 \right) = 1.
\label{eq:norm_uv}
\end{eqnarray}
Notice that in general the imaginary part of $\omega$ can be finite, signaling the dynamical instability of the static solution, and in such a case $(u_{{\boldsymbol i},n},v_{{\boldsymbol i},n})$ does not satisfy Eq.~(\ref{eq:norm_uv}).
However, in the systems that we consider here, $\omega$ is always real because the static solution is a ground state. 

Since one of our purposes in the present paper is to make a comparison between the results obtained by the analytical GL theory and the numerical Gutzwiller formalism in connection with the Higgs bound states, we need to relate $(u_{{\boldsymbol i},n},v_{{\boldsymbol i},n})$ with the fluctuations in the phase and amplitude of the superfluid order-parameter. Substituting
Eq.~(\ref{eq:fluc_f}) into Eq.~(\ref{eq:phi_f}) and linearizing the latter with respect to the fluctuations, we obtain
\begin{eqnarray}
\Phi_{\boldsymbol i} \simeq \tilde{\Phi}_{\boldsymbol i} 
+ \alpha_{\boldsymbol i} \cos\left(\omega t\right) 
+ i \tilde{\Phi}_{\boldsymbol i} \varphi_{\boldsymbol i} \sin\left(\omega t\right),
\end{eqnarray}
where
\begin{eqnarray}
\!\!\!\!\!\!\!\!\! \alpha_{\boldsymbol i} \!\!\!&=&\!\!\! \sum_{n}\! \left[ 
\tilde{f}_{{\boldsymbol i},n-1} \!
\left( u_{{\boldsymbol i},n} \! + \! v_{{\boldsymbol i},n} \right) 
\!+\!
 \tilde{f}_{{\boldsymbol i},n} \!
\left( u_{{\boldsymbol i},n-1} \! + \! v_{{\boldsymbol i},n-1} \right) 
\right]
\! ,
\label{eq:Gutz_alpha}
\\
\!\!\!\!\!\!\!\!\! \tilde{\Phi}_{\boldsymbol i}\varphi_{\boldsymbol i} \!\!\!&=&\!\!\!
\sum_{n}\! \left[ 
\tilde{f}_{{\boldsymbol i},n-1} \!
\left( v_{{\boldsymbol i},n} \! - \! u_{{\boldsymbol i},n} \right) 
\!+\!
 \tilde{f}_{{\boldsymbol i},n} \!
\left( u_{{\boldsymbol i},n-1} \! - \! v_{{\boldsymbol i},n-1} \right) 
\right]
\! .
\end{eqnarray}
$\alpha_{\boldsymbol i}$ and $\varphi_{\boldsymbol i}$ correspond to the phase and amplitude fluctuations.

\begin{figure}[tb]
     \includegraphics[scale=0.55]{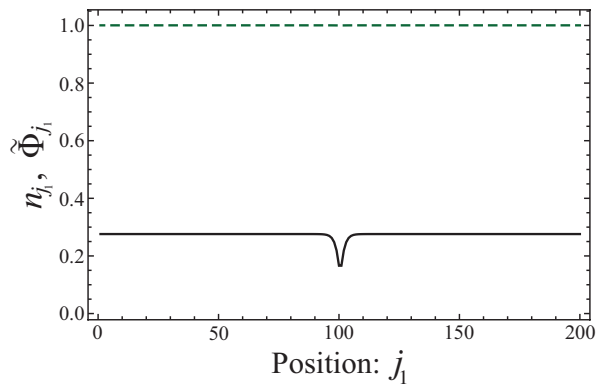}
   \caption{Spatial distributions of $n_{j_1}$ (green dashed line) and $\tilde{\Phi}_{j_1}$ (black solid line) for the ground state, where $zJ/U =0.18$ (or equivalently $(J-J_{\rm c})/J_{\rm c}=0.04912$), $\mu/U = \mu_{\rm c}/U$, $J'/J = 0.5$, $w=0$, and $L=200$.
   }
\label{fig:DensPsi}
\end{figure} 
\begin{figure}[tb]
     \includegraphics[scale=0.55]{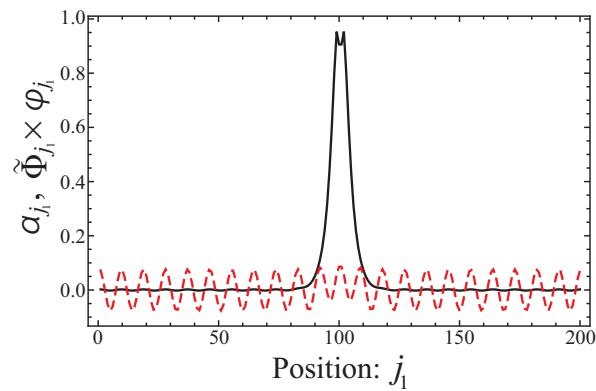}
   \caption{Spatial distributions of $\alpha_{j_1}$ (black solid line) and $\tilde{\Phi}_{j_1}\varphi_{j_1}$ (red dashed line) for a normal mode corresponding to a Higgs quasi-bound state with even parity, where $zJ/U =0.18$ (or equivalently $(J-J_{\rm c})/J_{\rm c}=0.04912$), $\mu/U = \mu_{\rm c}/U$, $J'/J = 0.5$, $w=0$, and $L=200$. The normal-mode frequency is given by $\hbar\omega = 0.1384 \, U$ while the bulk Higgs gap is $\Delta = 0.1568 \, U$.
   }
\label{fig:WF}
\end{figure} 

As an example of the computation using the above Gutzwiller formalism, we show the spatial distributions of the density $n_{j_1} = \langle \hat{n}_{j_1} \rangle$ and the order-parameter amplitude $\tilde{\Phi}_{j_1}$ for a ground state in Fig.~\ref{fig:DensPsi} and those of the amplitude and phase fluctuations, $\alpha_{j_1}$ and $\tilde{\Phi}_{j_1} \times \varphi_{j_1}$, for a normal mode corresponding to a Higgs bound state with even parity in Fig.~\ref{fig:WF}. There we set $zJ/U =0.18$ (or equivalently $(J-J_{\rm c})/J_{\rm c}=0.04912$), $\mu/U = \mu_{\rm c}/U $, $J'/J=0.5$, and $L=200$, where $(zJ_{\rm c}/U, \mu_{\rm c}/U)=(3-2\sqrt{2}, \sqrt{2}-1)$ denotes the critical point of the Mott transition at unit filling in the $(zJ/U,\mu/U)$ plane within the mean-field approximation~\cite{vanoosten-01}. In Fig.~\ref{fig:DensPsi}, we see that while $\tilde{\Phi}_{j_1}$ clearly diminishes near the potential barrier, $n_{j_1}$ remains almost constant in space. This manifests the approximate particle-hole symmetry emerging in a vicinity of the critical point. In other words, both particle- and hole-fluctuations contribute equally to the formation of the superfluid order-parameter such that the spatial modulation of the order-parameter amplitude does not lead to that of the density.

\begin{figure}[tb]
     \includegraphics[scale=0.6]{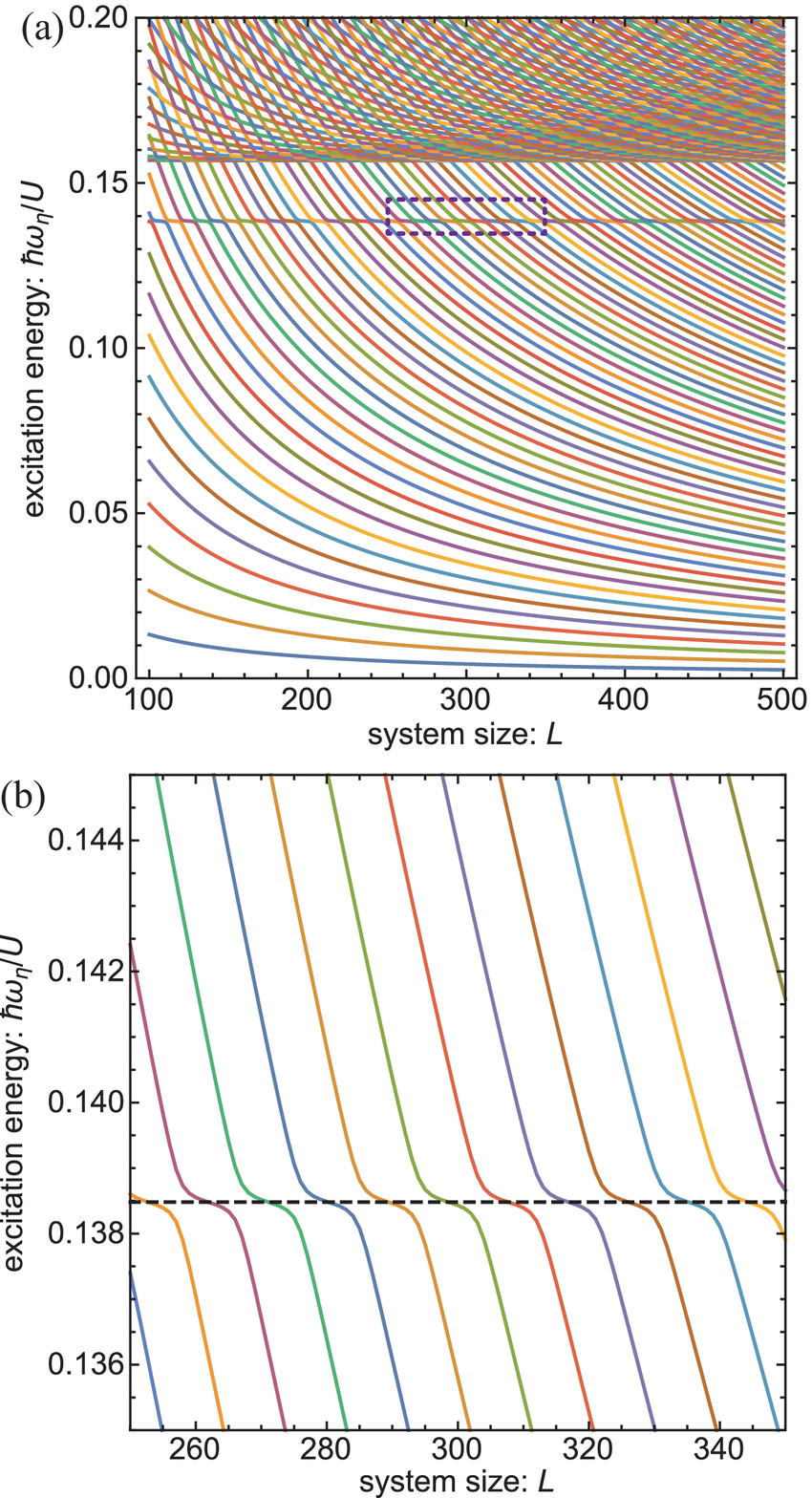}
   \caption{Excitation energies of the normal modes with even parity as functions of the system size $L$, where $zJ/U =0.18$ (or equivalently $(J-J_{\rm c})/J_{\rm c}=0.04912$), $\mu/U = \mu_{\rm c}/U$, $J'/J = 0.5$, and $w=0$. (b) is a magnified view of the region indicated by the dotted square in (a). The dashed line in (b) marks the resonance energy $\hbar \omega_{\rm res} = 0.1385 U$.   }
\label{fig:Espe}
\end{figure} 

In Fig.~\ref{fig:WF}, we see that $\alpha_{j_1}$ is localized around the position of the potential barrier and this property is a clear signature of the Higgs bound state. However, the phase fluctuation delocalized in the entire system is small but finite, and such mixing with the delocalized mode means that this state is not a true bound state but a quasi-bound state, in which the initial localized state decays into the coupled delocalized mode at finite time. In contrast, the Higgs bound state solution shown in Sec.~\ref{subsec:GL} is purely an amplitude mode because $\mu/U = \mu_{\rm c}/U$ implies $K_0 = 0$, i.e., the time-dependent GL equation (\ref{eq:GL}) is particle-hole symmetric. Hence, this slight mixing of the Higgs mode with the NG mode is due to higher order corrections breaking the particle-hole symmetry, such as the third-order time derivative term $\frac{\partial^3 \Phi}{\partial t^3}$ that is proportional to $\omega^3$.

\begin{figure*}[tb]
     \includegraphics[scale=0.43]{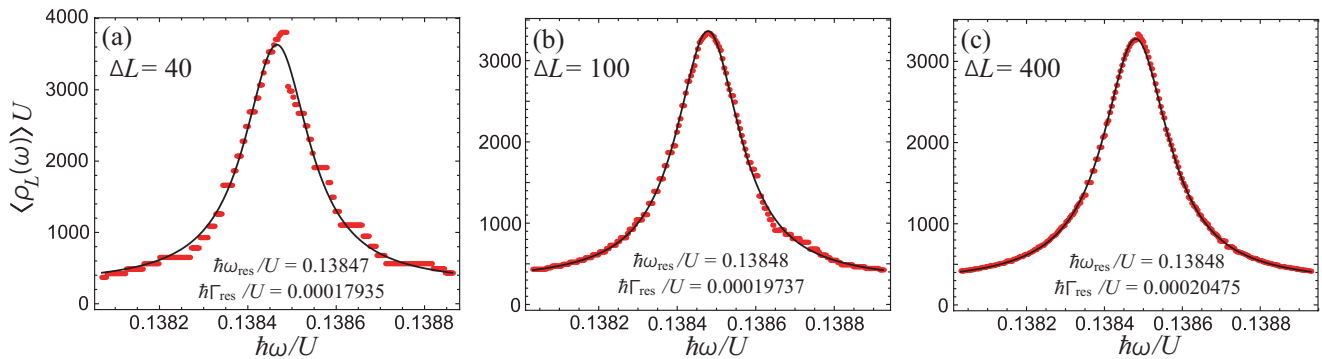}
   \caption{Averaged density of states $\langle \rho_L(\omega) \rangle$ for the normal modes with even parity at several values of $\Delta L$, where $L=300$, $zJ/U =0.18$ (or equivalently $(J-J_{\rm c})/J_{\rm c}=0.04912$), $\mu/U = \mu_{\rm c}/U$, $J'/J = 0.5$, and $w=0$. The black solid line represents the best fit to the data with the fitting function of Eq.~(\ref{eq:Lorentz}).}
\label{fig:rho_delL}
\end{figure*} 
\begin{figure*}[tb]
     \includegraphics[scale=0.43]{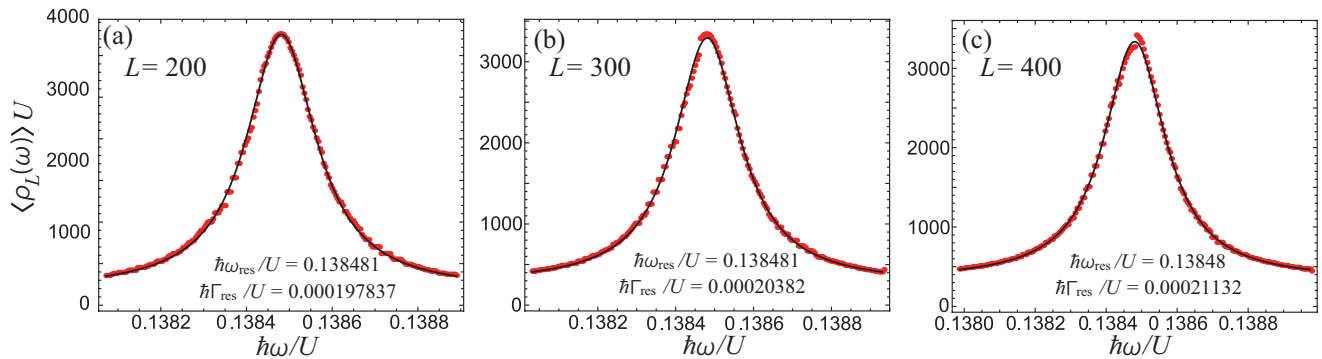}
   \caption{Averaged density of states $\langle \rho_L(\omega) \rangle$ for the normal modes with even parity at several values of $L$, where $\Delta L = 200$, $zJ/U =0.18$ (or equivalently $(J-J_{\rm c})/J_{\rm c}=0.04912$), $\mu/U = \mu_{\rm c}/U$, $J'/J = 0.5$, and $w=0$. The black solid line represents the best fit to the data with the fitting function of Eq.~(\ref{eq:Lorentz}).}
\label{fig:rho_L}
\end{figure*} 

%
\subsection{Stabilization method}
\label{subsec:SM}

In Fig.~\ref{fig:Espe}, we show the excitation energies of the normal modes with even parity versus the system size $L$ computed by the Gutzwiller method. When $L$ increases, most of the energies decay like $\sim O(L^{-1})$ or $\sim \Delta + O(L^{-2})$, corresponding to delocalized NG or Higgs modes. In contrast, there is a mode whose excitation energy is almost constant with increasing $L$. This indicates that the mode is localized at a short range. Looking at the phase and amplitude fluctuations shown in Fig.~\ref{fig:WF} for specific $L$($=200$), we indeed see that the amplitude fluctuation is localized around the potential barrier. However, in Fig.~\ref{fig:Espe}(b) where a magnified view of the excitation energy of the localized mode is depicted, we see that the energy of this mode slightly depends on $L$ by making several avoided crossings with delocalized NG modes. In other words, the localized state is not a true bound state but a quasi-bound state as was discussed in the previous section. In this section, we explain a stabilization method~\cite{mandelshtam-93} for determining the resonance energy $\omega_{\rm res}$ and the line width $\Gamma$ of the localized mode, which are independent of $L$.

Let us write symbolically the density of states  for normal modes $\rho(\omega)$ as
\begin{eqnarray}
\rho(\omega) = \rho^{Q}(\omega) +  \rho^{P}(\omega),
\end{eqnarray}
where $\rho^{Q}(\omega)$ and $\rho^{P}(\omega)$ represent the contribution from the quasi-bound state and that from the delocalized modes in the background.
In the stabilization method, we utilize the general facts that the density of quasi-bound states takes the following Lorentzian form \cite{mandelshtam-93},
\begin{eqnarray}
\rho^{Q}(\omega) \propto \frac{1}{(\omega_{\rm res}- \omega)^2 + \Gamma^2/4},
\end{eqnarray}
and that the Lorentzian peak is more pronounced than $\rho^{P}(\omega)$ in the background. In other words, once $\rho(\omega)$ is numerically given, one can obtain $\omega_{\rm res}$
and $\Gamma$ through a Lorentzian fit. 

To construct $\rho(\omega)$ numerically, we start with the density of states at a fixed value of $L$,
\begin{eqnarray}
\rho_L(\omega) = \sum_{\zeta} \delta(\omega_{\zeta}(L) - \omega).
\end{eqnarray}
where $\omega_{\zeta}(L)$ denotes the eigenvalue of Eqs.~(\ref{eq:bogo_u}) and (\ref{eq:bogo_v}) with quantum number $\zeta$.
Since $\rho^{Q}(\omega)$, which we are interested, is expected to be independent of $L$ at sufficiently large $L$ compared with $\xi$, averaging $\rho_L(\omega)$ with respect to $L$ should not spoil the Lorentzian peak stemming from $\rho^{Q}(\omega)$. The average of $\rho_L(\omega)$ is defined as
\begin{eqnarray}
\langle \rho_L(\omega) \rangle = \frac{1}{\Delta L} \int_{L-\Delta L/2}^{L+\Delta L/2} d\tilde{L}
\, \rho_{\tilde{L}}(\omega).
\label{eq:rho_ave}
\end{eqnarray}
Using a well-known formula,
\begin{eqnarray}
\int dx\, \delta\left(f_0 - f(x)\right) = \left| \frac{df}{dx} \right|_{f(x)=f_0},
\end{eqnarray}
Eq.~(\ref{eq:rho_ave}) can be rewritten as
\begin{eqnarray}
\langle \rho_L(\omega) \rangle = \frac{1}{\Delta L} \sum_{\zeta} \left(\left| \frac{d \omega_{\zeta}(\tilde{L})}{d\tilde{L}} \right|_{\omega_{\zeta}(\tilde{L})=\omega}\right)^{-1}.
\label{eq:rho_final}
\end{eqnarray}
The summation with respect to $\zeta$ in the right-hand side of Eq.~(\ref{eq:rho_final}) is taken only for $\tilde{L}$ satisfying $L-\Delta L/2 \leq \tilde{L}\leq L+ \Delta L/2$.
The formula of Eq.~(\ref{eq:rho_final}) converts the $L$-dependence of the excitation energies, which is shown in Fig.~\ref{fig:Espe}, to the averaged density of states.
However, since we are working in a discrete system on a lattice, we need to replace the condition $\omega_{\zeta}(\tilde{L})=\omega$ with $\omega_{\zeta}(\tilde{L}+1)<\omega<\omega_{\zeta}(\tilde{L})$. Moreover, the derivative by $\tilde{L}$ in the discrete system is defined as
\begin{eqnarray}
 \frac{d \omega_{\zeta}(\tilde{L})}{d\tilde{L}} = 
 \omega_{\zeta}(\tilde{L}+1)- \omega_{\zeta}(\tilde{L}).
\end{eqnarray}
With these slight modifications, we can apply the formula of Eq.~(\ref{eq:rho_final}) for evaluating the density of states for normal modes in our discrete Bose-Hubbard system.

In Fig.~\ref{fig:rho_delL}, we show $\langle \rho_L(\omega) \rangle$ by varying $\Delta L$ at fixed $L$($=300$). In order to extract the resonance energy $\omega_{\rm res}$ and the line width $\Gamma$ from $\langle \rho_L(\omega) \rangle$, we use the following fitting function,
\begin{eqnarray}
f(x) = \frac{A}{(B - x)^2 + C^2/4} + D,
\label{eq:Lorentz}
\end{eqnarray}
where $A$, $B$, $C$, and $D$ are free parameters. The extracted values of $B$ and $C$ correspond to $\omega_{\rm res}$ and $\Gamma$, respectively. In Fig.~\ref{fig:rho_delL}, we see that when $\Delta L$ increases, $\langle \rho_L(\omega) \rangle$ becomes smoother and its shape approaches the Lorentzian form. We also show the $L$-dependence of $\langle \rho_L(\omega) \rangle$ for fixed $\Delta L$ in Fig.~\ref{fig:rho_L}. We see that the result at $L=200$ and $\Delta L = 200$ is already converged up to the fifth place of decimals. Hence, we will take $L=200$ and $\Delta L = 200$ for the calculations shown below as long as $zJ/U\geq 0.18$. On the other hand, when $zJ/U< 0.18$, we will take $L=300$ and $\Delta L = 400$ for the following reason. When $zJ/U$ decreases and approaches the critical value $zJ_{\rm c}/U$, $\omega_{\rm res}$ decreases and approaches zero such that the density of states for delocalized NG modes near $\omega = \omega_{\rm res}$ also decreases. This means that at fixed $L$ and $\Delta L$ 
the number of states contributed to the summation in Eq.~(\ref{eq:rho_final}) also decreases. Hence, in order to keep a sufficient number of samples in the averaging, the closer the system is to the critical point, the larger we need to make $L$ and $\Delta L$.

\section{Results}
\label{sec:result}
%
\subsection{Comparison between GL and Gutzwiller results}
\label{subsec:comp}
%
\begin{figure}[tb]
     \includegraphics[scale=0.55]{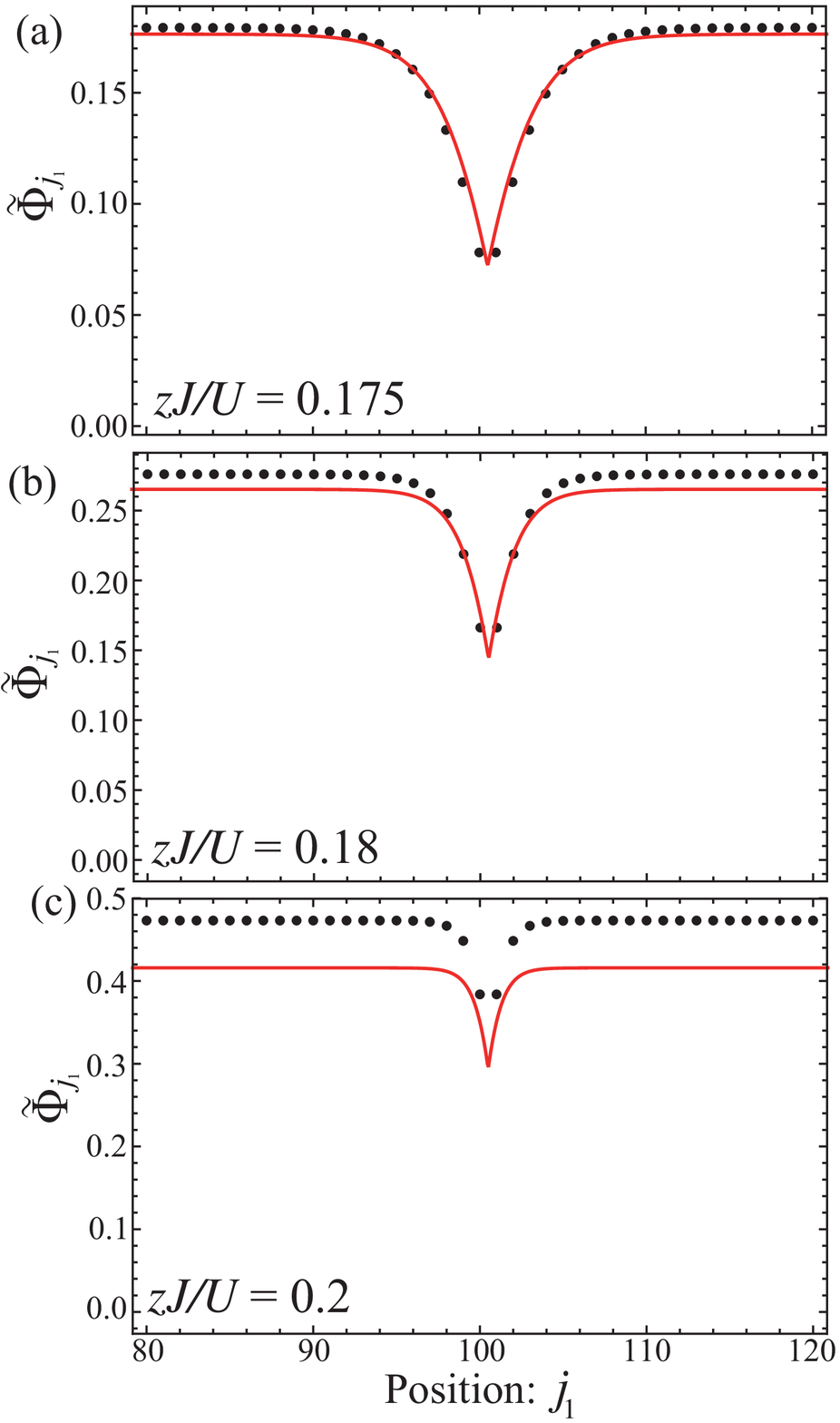}
   \caption{Comparison between the results by the GL and Gutzwiller methods regarding the spatial distribution of the static order parameter $\tilde{\Phi}_{j_1}$ in the ground state, where $L=200$, $\mu/U = \mu_{\rm c}/U$, $J'/J = 0.5$, and $w=0$. $zJ/U=0.175$, $0.18$, and $0.2$ imply $(J-J_{\rm c})/J_{\rm c}=0.01997$, $0.04912$, and $0.1657$. The red solid line and the black dots represent the GL and Gutzwiller results. 
   }
\label{fig:PsiComp}
\end{figure} 
\begin{figure}[tb]
     \includegraphics[scale=0.55]{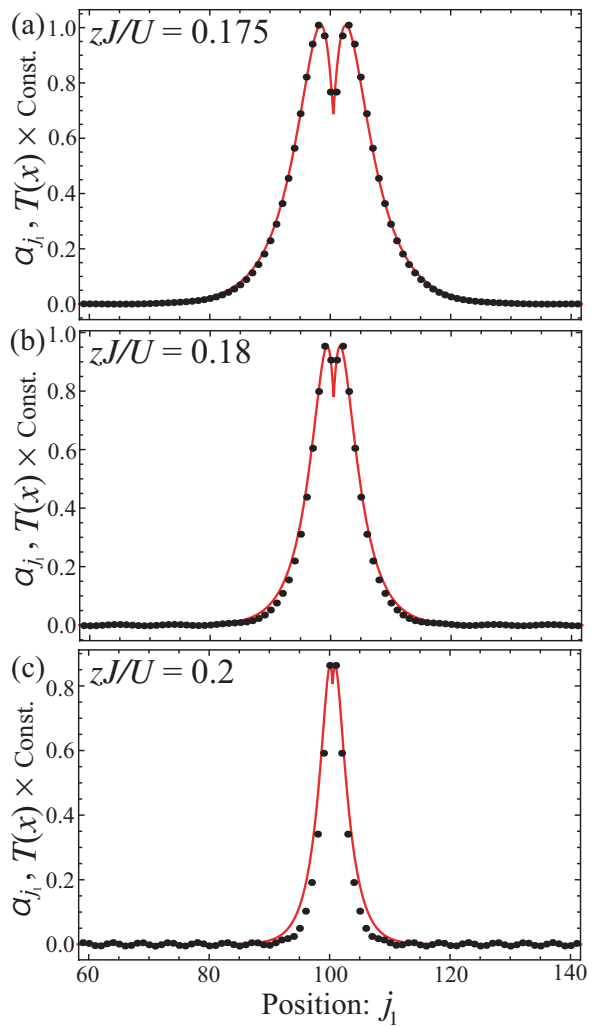}
   \caption{Comparison between the Gutzwiller and GL regarding the spatial distribution of the amplitude fluctuation from the ground state in the Higgs quasi-bound state with even parity, where $L=200$, $\mu/U = \mu_{\rm c}/U$, $J'/J = 0.5$, and $w=0$. $zJ/U=0.175$, $0.18$, and $0.2$ imply $(J-J_{\rm c})/J_{\rm c}=0.01997$, $0.04912$, and $0.1657$. The red solid line and the black dots represent the GL and Gutzwiller results, namely $T(x)$ of Eq.~(\ref{eq:Tx}) and $\alpha_{j_1}$ of Eq.~(\ref{eq:Gutz_alpha}). We multiply $T(x)$ with a constant such that the maximum value of $T(x)$ becomes equal to that of $\alpha_{j_1}$.
   }
\label{fig:AmpComp}
\end{figure} 

In Figs.~\ref{fig:PsiComp} and \ref{fig:AmpComp}, we show the spatial distribution of the static superfluid order-parameter for the ground state and that of the amplitude fluctuation from the ground state. There we compare the results by the Gutzwiller approach (black dots) to those by the GL theory (red solid lines). We clearly see that when $zJ/U$ approaches its critical value, the GL results approach the Gutzwiller results. The agreement is quantitative already at $zJ/U=0.18$. These quantitative comparisons corroborate that the Gutzwiller approach can correctly capture the physics of Higgs bound states.

\subsection{Varying $zJ/U$}
\label{subsec:J}
%
\begin{figure}[tb]
     \includegraphics[scale=0.6]{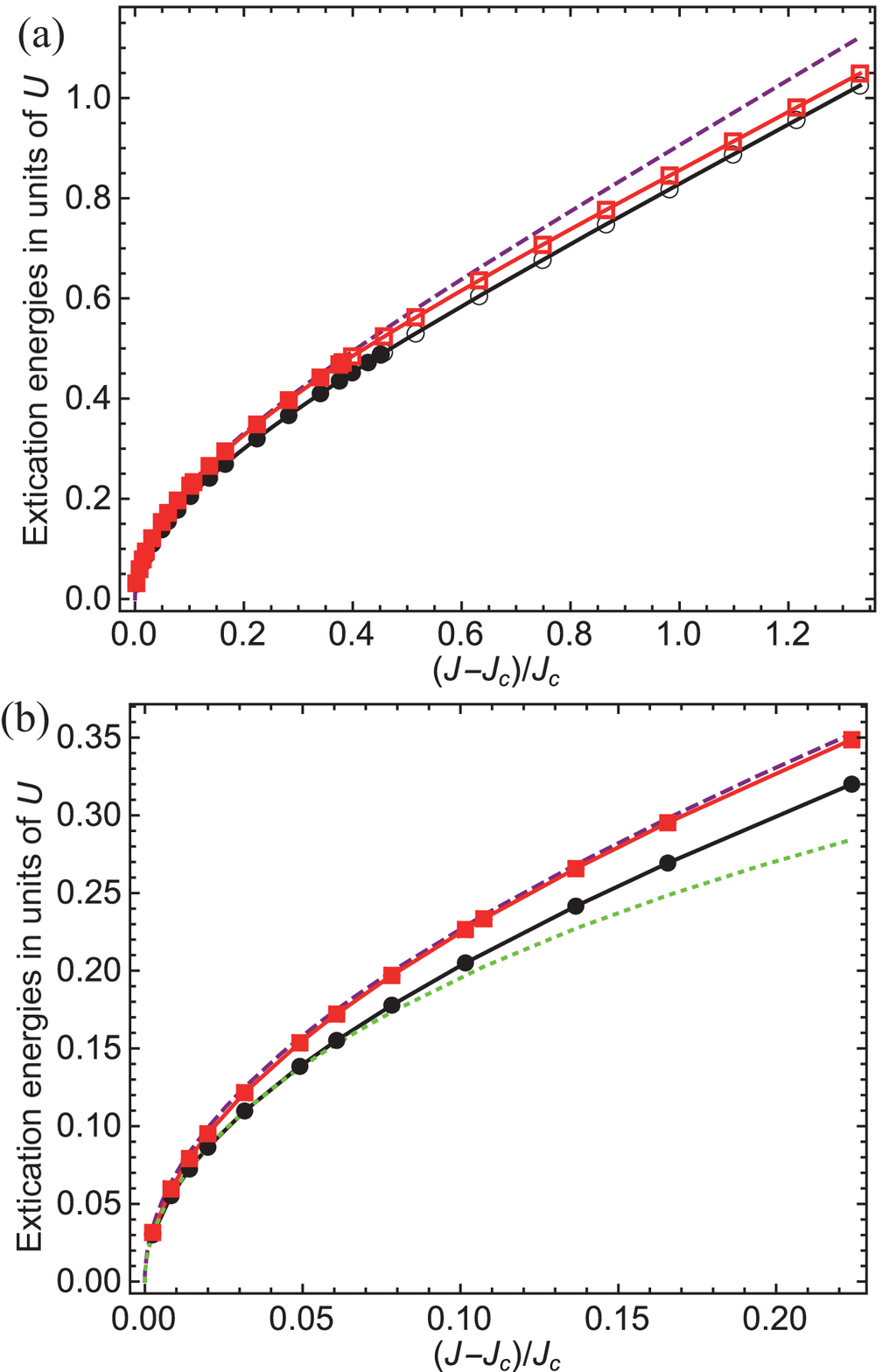}
   \caption{Binding energies as functions of $(J-J_{\rm c})/J_{\rm c}$ for $\mu/U =\mu_{\rm c}/U$, $J'/J = 0$ and $w=0$. The black circles (red squares) represent the energy of the Higgs bound state with even (odd) parity calculated by the Gutzwiller approach. The filled symbols represent the resonance energy $\hbar\omega_{\rm res}$ of quasi-bound states computed by the stabilization method while the open symbols represent the excitation energy of true bound states at $L=200$. The solid lines are guides to the eye. The purple dashed line represents the gap of the delocalized Higgs mode $\Delta$. In (b), where a magnified view of (a) around the critical point is depicted, the green dotted line represents the energy of the Higgs bound state with even parity evaluated from the GL theory. 
   }
\label{fig:Ew0}
\end{figure} 
\begin{figure}[tb]
\includegraphics[scale=0.6]{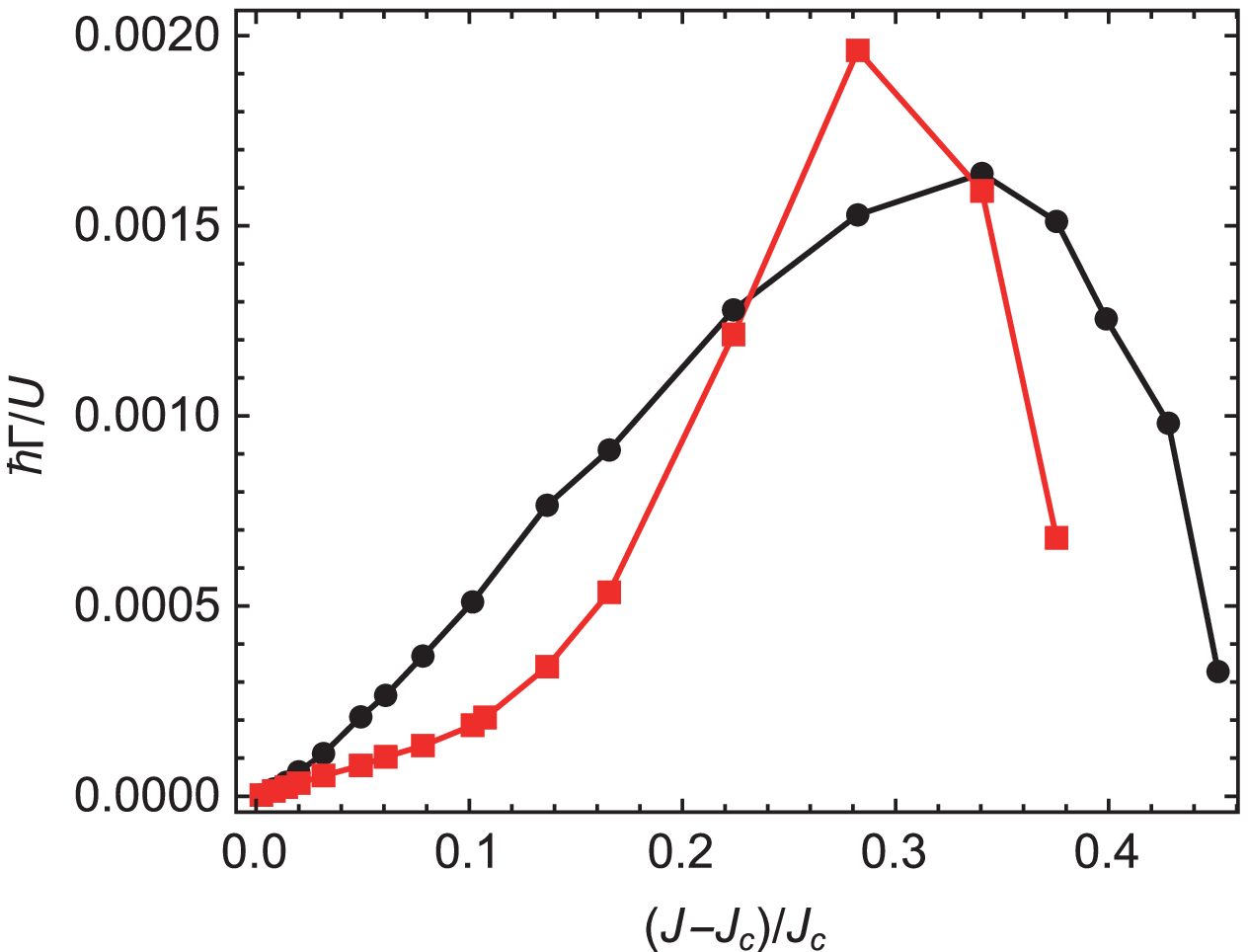}
\caption{
Line widths of the resonant states $\Gamma$ as functions of $(J-J_{\rm c})/J_{\rm c}$ for $\mu/U=\mu_{\rm c}/U$, $J'/J=0.5$, and $w=0$. The black circles (red squares) represent the width of the Higgs bound state with even (odd) parity. The solid lines are guides to the eye.
}
\label{fig:Gw0}
\end{figure}

In Fig.~\ref{fig:Ew0}, we show the binding energies versus $(J-J_{\rm c})/J_{\rm c}$ for the Higgs bound states with even and odd parities along the line of $\mu/U = \mu_{\rm c}/U$, which corresponds to $K_0=0$. When $(J-J_{\rm c})/J_{\rm c}$ increases, the binding energies for both states monotonically increases.  As for the even bound state, we compare the Gutzwiller result with the GL result, which is plotted by the green dotted line in Fig.~\ref{fig:Ew0}(b). At $(J-J_{\rm c})/J_{\rm c}<0.05$, they precisely agree. They start to deviate noticeably around $(J-J_{\rm c})/J_{\rm c}=0.05$ and the deviation becomes larger when $(J-J_{\rm c})/J_{\rm c}$ increases. Nevertheless, our Gutzwiller analysis clearly shows that the (quasi-)bound states survive even in a parameter region where the GL theory completely fails.

In Fig.~\ref{fig:Gw0}, we show the line width $\Gamma$ versus $(J-J_{\rm c})/J_{\rm c}$ for the even and odd Higgs bound states along the line of $\mu/U = \mu_{\rm c}/U$. When $(J-J_{\rm c})/J_{\rm c}$ increases near zero, $\Gamma$ also increases from zero. However, $\Gamma$ remains much smaller than $\omega_{\rm res}$ so that the resonance can be identified as a sharp peak in the density of states. $\Gamma$ is maximized around $(J-J_{\rm c})/J_{\rm c}=0.3$ and it approaches zero when $(J-J_{\rm c})/J_{\rm c}$ increases further. This indicates that the quasi-bound states with finite life time turn into true bound states for large $(J-J_{\rm c})/J_{\rm c}$. 

\begin{figure}[tb]
     \includegraphics[scale=0.55]{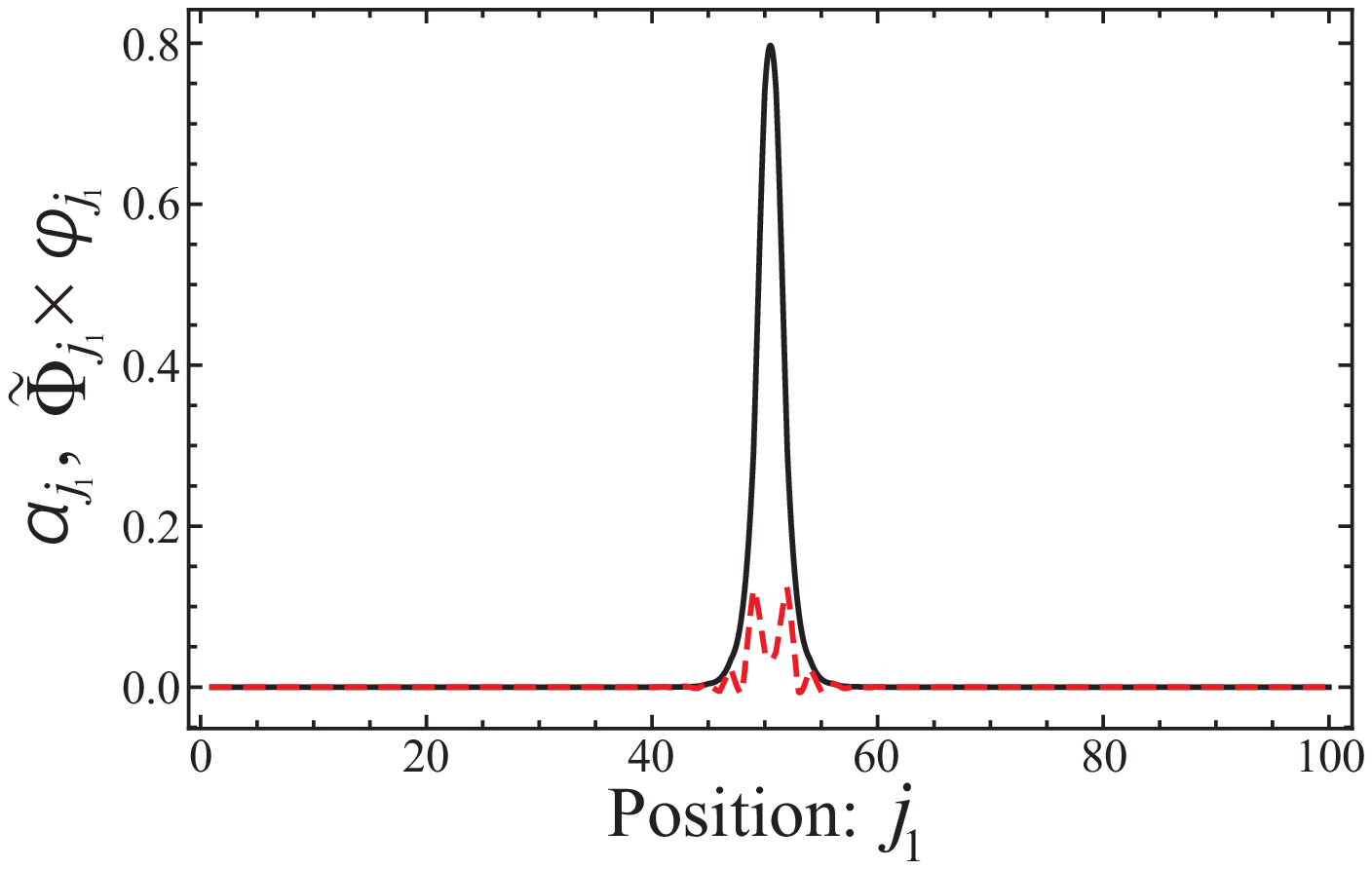}
   \caption{Spatial distributions of $\alpha_{j_1}$ (black solid line) and $\tilde{\Phi}_{j_1}\varphi_{j_1}$ (red dashed line) for the Higgs bound state with even parity, where $zJ/U =0.26$ (or equivalently $(J-J_{\rm c})/J_{\rm c}=0.5154$), $\mu/U = \mu_{\rm c}/U$, $J'/J = 0.5$, $w=0$, and $L=100$.
   }
\label{fig:WF2}
\end{figure} 
\begin{figure}[tb]
\includegraphics[scale=0.6]{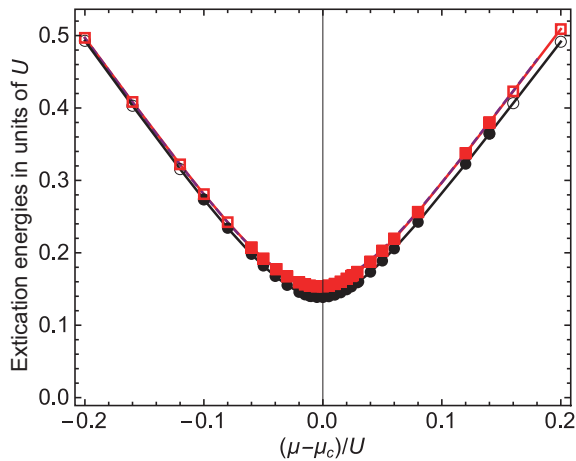}
\caption{
Binding energies as functions of $(\mu -\mu_{\rm c})/U$ for $zJ/U=0.18$, $J'/J = 0.5$ and $w=0$. The black circles (red squares) represent the energy of the bound state with even (odd) parity calculated by the Gutzwiller approach. The filled symbols represent the resonance energy of quasi-bound states computed by the stabilization method while the open symbols represent the excitation energy of true bound states at $L=100$. The solid lines are guides to the eye. The purple dashed line represents the gap $\Delta$ of the normal mode that becomes the delocalized Higgs mode at $\mu/U \simeq \mu_{\rm c}/U$.
}
\label{fig:Ew0mu}
\end{figure}

\begin{figure}[tb]
\includegraphics[scale=0.6]{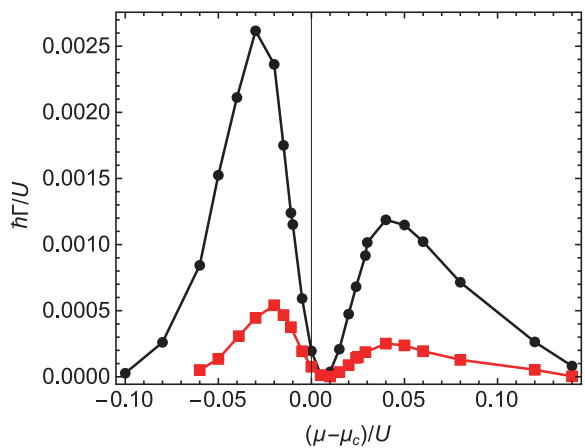}
\caption{
Line widths of the resonant states $\Gamma$ as functions of $(\mu -\mu_{\rm c})/U$ for $zJ/U=0.18$ (or equivalently $(J-J_{\rm c})/J_{\rm c}=0.04912$),  and $w=0$. The black circles (red squares) represent the width of the Higgs bound state with even (odd) parity. The solid lines are guides to the eye. 
}
\label{fig:Gw0mu}
\end{figure}
\begin{figure}[tb]
     \includegraphics[scale=0.55]{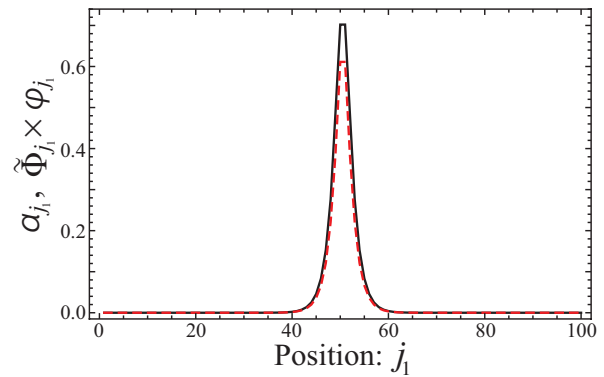}
   \caption{Spatial distributions of $\alpha_{j_1}$ (black solid line) and $\tilde{\Phi}_{j_1}\varphi_{j_1}$ (red dashed line) for a normal mode corresponding to the bound state with even parity, where $zJ/U =0.18$ (or equivalently $(J-J_{\rm c})/J_{\rm c}=0.04912$), $\mu/U = \mu_{\rm c}/U + 0.2$, $J'/J = 0.5$, $w=0$, and $L=100$. 
   }
\label{fig:WF3}
\end{figure} 
\begin{figure}[b]
\includegraphics[scale=0.6]{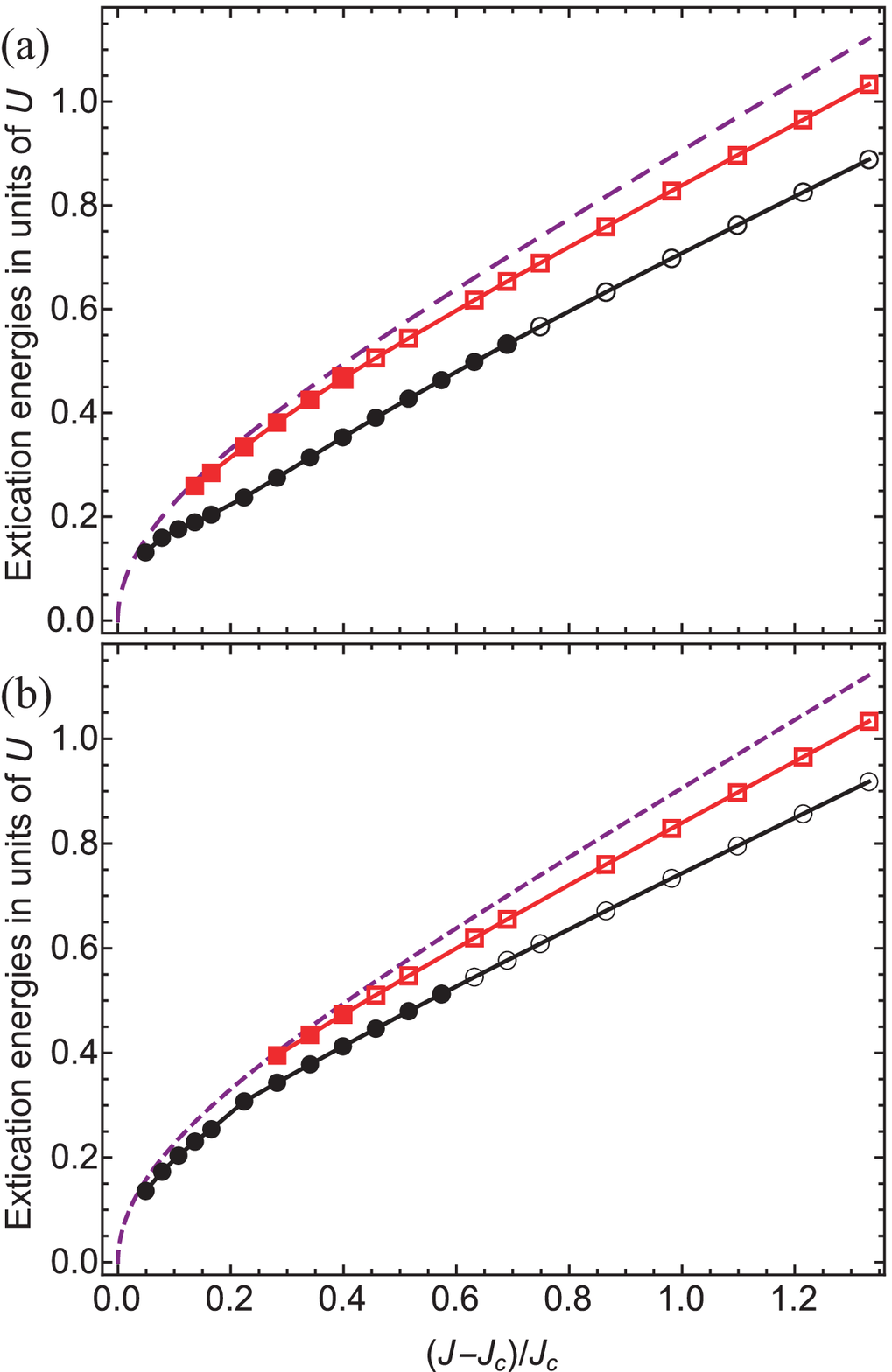}
\caption{\label{fig:Ew1}
Binding energies as functions of $(J-J_{\rm c})/J_{\rm c}$ for $\mu/U =\mu_{\rm c}/U$, $J'/J = 0$ and $w=1$. In (a) and (b), the binding energies for the even- and odd-parity states calculated by the Gutzwiller approach are shown. The filled symbols represent the resonance energy $\hbar\omega_{\rm res}$ of quasi-bound states computed by the stabilization method while the open symbols represent the excitation energy of true bound states at $L=200$. The solid lines are guides to the eye. The purple dashed line represents the gap of the delocalized Higgs mode $\Delta$. 
}
\end{figure}

In Fig.~\ref{fig:WF2}, we show the spatial distributions of the amplitude and phase fluctuations for the Higgs bound state with even parity at $(J-J_{\rm c})/J_{\rm c}=0.5154$. In contrast to Fig.~\ref{fig:WF}, where the system is much closer to the critical point as $(J-J_{\rm c})/J_{\rm c}=0.04912$, we clearly see that the phase fluctuation is also localized around the potential barrier. Nevertheless, since the amplitude fluctuation dominates over the phase fluctuation, we still call this bound state as a Higgs mode. In Fig.~\ref{fig:Ew0}(a), the excitation energies of the true bound states are plotted by the open symbols. 
It is smoothly connected with the resonance energy of the quasi-bound state.

The physical origin of the transition from quasi- to true bound state is rather simple. When $(J-J_{\rm c})/J_{\rm c}$ increases, the resonance energy of the quasi-bound state increases and it exceeds the maximum energy of the branch of the delocalized NG mode above a certain threshold value of $(J-J_{\rm c})/J_{\rm c}$. Since there is no delocalized mode with the energy equal to the resonance energy, the bound state can not decay into delocalized modes and it stands as a true bound state.

It is worth emphasizing that the transition from quasi- to true bound state is an artifact of the Gutzwiller approach and should not be observed in exact theoretical analyses of the 3D Bose-Hubbard model or in experiments. Specifically, we neglect effects of interactions between collective modes and quantum fluctuations in the Gutzwiller approximation. These effects allow Higgs modes to decay into multiple NG modes and such decay channels should exist for the Higgs bound states, thus making its life time finite.

\begin{figure}[tb]
\includegraphics[scale=0.6]{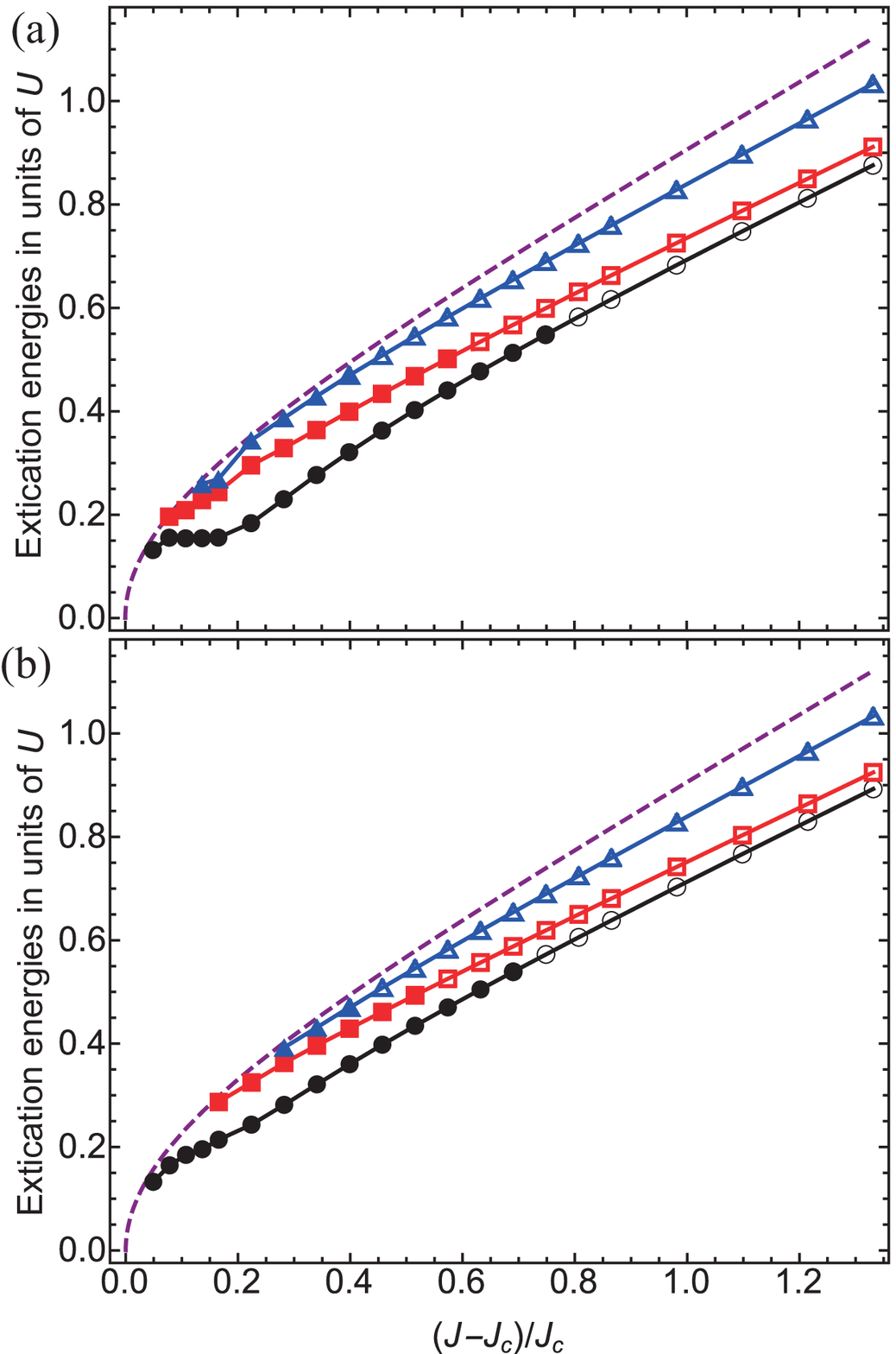}
\caption{\label{fig:Ew2}
Binding energies as functions of $(J-J_{\rm c})/J_{\rm c}$ for $\mu/U =\mu_{\rm c}/U$, $J'/J = 0$ and $w=2$. In (a) and (b), the binding energies for the even- and odd-parity states calculated by the Gutzwiller approach are shown. The filled symbols represent the resonance energy $\hbar\omega_{\rm res}$ of quasi-bound states computed by the stabilization method while the open symbols represent the excitation energy of true bound states at $L=200$. The solid lines are guides to the eye. The purple dashed line represents the gap of the delocalized Higgs mode $\Delta$. 
}
\end{figure}
%
\subsection{Varying $\mu/U$}
In Figs.~\ref{fig:Ew0mu} and \ref{fig:Gw0mu}, we show the binding energies and the line widths for the bound states versus $(\mu - \mu_{\rm c})/U$ at $zJ/U =0.18$. In Fig.~\ref{fig:Ew0mu}, we see that the binding energies are minimized at $(\mu - \mu_{\rm c})/U \simeq 0$. When $|\mu - \mu_{\rm c}|/U$ increases and exceeds certain threshold values, $\Gamma$ becomes zero as shown in Fig.~\ref{fig:Gw0mu}. This means that the quasi-bound states turn into true bound states for the reason discussed in Sec.~\ref{subsec:J}. The increase in $(\mu - \mu_{\rm c})/U$ corresponds to the increase in $|K_0|$ such that in a parameter region far apart from the line of $\mu/U = \mu_{\rm c}/U$, the bound states are not dominated by amplitude fluctuations.  In other words, they can not be interpreted as a Higgs amplitude mode but as a mode corresponding to a single-particle (or hole) excitation. For instance, in Fig.~\ref{fig:WF3}, we show the spatial distributions of the amplitude and phase fluctuations at $(\mu - \mu_{\rm c})/U = 0.2 $, where the contributions from the two kinds of fluctuation are indeed comparable. This implies that the particle-density fluctuation dominates this mode.

Within the GL theory shown in Sec.~\ref{subsec:GL}, at $K_0 = 0$, which corresponds to $\mu/U = \mu_{\rm c}/U$, there is no coupling between the NG and Higgs modes such that the line width $\Gamma$ is minimized and equal to zero at $\mu/U = \mu_{\rm c}/U$. However, in Fig.~\ref{fig:Gw0mu}, the value of $(\mu - \mu_{\rm c})/U$, which gives minimum $\Gamma$, is slightly shifted from zero. The shifted minimum can be interpreted as the point where the $K_0$ term cancels the higher order corrections, which breaks the particle-hole symmetry, to the GL equation.

\subsection{Varying $w$}
When $w=0$, the binding energies of the Higgs bound states are rather close to the bulk Higgs gap $\Delta$. More specifically, that of the even bound state is roughly 90$\%$ of $\Delta$. This property is disadvantageous for observing the Higgs bound states as a separate resonance peak in a response to external perturbations, because in realistic systems the peak should be broadened due to thermal and quantum fluctuations neglected in the Gutzwiller approximation. In the following, we suggest that the energy separation from the lowest Higgs bound state can be enlarged by using the tunability of the barrier width.

In Figs.~\ref{fig:Ew1} and \ref{fig:Ew2}, we show the binding energies for the Higgs bound states versus $(J-J_{\rm c})/J_{\rm c}$ at $w=1$ and $2$. There we clearly see that when $w$ increases, the number of the (quasi-)bound states increases. This is simply because a wider barrier provides with space for accommodating the wave functions with more nodes. Moreover, at an optimal value of $(J-J_{\rm c})/J_{\rm c}$, the separation of the lowest binding energy from those of the other even-parity modes becomes remarkably larger at $w=2$. Such separation makes it easier to observe a resonance peak corresponding to the lowest Higgs bound state in experiment.

\section{Conclusions}
\label{sec:conc}
We have studied collective-mode properties of a Bose-Hubbard system in the presence of local suppression of the hopping amplitude, which acts for the superfluid order-parameter as a potential barrier preserving the particle hole symmetry. Specifically, we analyze localized (bound) states of Higgs amplitude mode with use of the Guzwiller mean-field approximation combined with a stabilization method. The employed method is advantageous over the Ginzburg-Landau (GL) theory used in the previous work~\cite{nakayama-15} in the sense that the former is applicable to a much broader parameter region. The agreement between the results obtained by the Gutzwiller and GL methods near the Mott transition corroborated that the former method can properly capture the physics of the Higgs bound states. We showed that the Higgs bound states survive even in a parameter region where the GL theory is invalid. Moreover, it was also shown that when the width of the potential barrier increases, the excitation energy of the lowest Higgs bound state decreases and that it can be well separated from the energies of other collective modes. These properties facilitate experimental observation of the Higgs bound state.

We found the transition from the quasi-bound state to a true bound state. However, we argued that the emergence of the true bound state should be an artifact of the Gutzwiller approximation such that it will not be observed in exact numerical analyses of the Bose-Hubbard model or in experiments. 

Throughout the present paper, we focused on a 3D system, in which the Gutzwiller and GL methods are reliable at least at a qualitative level. However, the most advanced experimental studies on Higgs modes in ultracold-gas systems have been performed in 2D optical-lattice systems~\cite{endres-12}, where quantum-gas microscope techniques with single-site resolution were used. Hence, it will be interesting to investigate effects of quantum and thermal fluctuations on the Higgs bound states at two dimension, especially whether they can be present as well-defined collective modes exhibiting a sharp peak in a response function, e.g., by means of quantum Monte Carlo simulations.

\begin{acknowledgments}
The authors thank T.~Nakayama and T.~Nikuni for useful discussions and collaboration on related work. This work was supported by KAKENHI from Japan Society for the Promotion of Science: Grants No.~25220711 (I.D.) and CREST, JST No. JPMJCR1673.
\end{acknowledgments}


\end{document}